\documentclass{article}

\usepackage[preprint]{neurips_2021}

\usepackage{float}
\usepackage{comment}
\usepackage[utf8]{inputenc} 
\usepackage[T1]{fontenc}    

\usepackage{url}            
\usepackage{booktabs}       
\usepackage{amsfonts}       
\usepackage{nicefrac}       
\usepackage{microtype}      
\usepackage{xcolor}         
\usepackage{amsmath,amssymb}
\usepackage{pifont}
\newcommand{\cmark}{\color{green}{\ding{51}}}%
\newcommand{\xmark}{\color{red}{\ding{55}}}%
\usepackage{makecell}

\usepackage[draft]{minted} 

\usepackage{graphicx} 
\usepackage{booktabs} 
\usepackage{bbding} 

\newcommand{\DP}{$\partial$P}
\newcommand{\AD}{AbstractDifferentiation.jl}

\usepackage{hyperref}       

\title{\AD: Backend-Agnostic Differentiable Programming in Julia}

\author{%
  Frank~Schäfer \\
  Department of Physics \\ 
  University of Basel, Switzerland \\
  \texttt{frank.schaefer@unibas.ch} \\
  \And
  Mohamed~Tarek \\
  Pumas-AI Inc., USA \\
  UNSW Canberra, Australia \\
  \texttt{mohamed@pumas.ai} \\
  \And
  Lyndon~White \\
  Invenia~Labs\\
  Cambridge, UK\\
  \texttt{lyndon.white@invenialabs.co.uk}\\
  \And
  Chris~Rackauckas \\
  Massachusetts Institute of Technology, USA\\
  Julia Computing Inc., USA\\
  Pumas-AI Inc., USA\\
  \texttt{crackauc@mit.edu}\\
}

\begin{document}

\maketitle

\begin{abstract}
  No single Automatic Differentiation (AD) system is the optimal choice for all problems. This means informed selection of an AD system and combinations can be a problem-specific variable that can greatly impact performance. In the Julia programming language, the major AD systems target the same input and thus in theory can compose. Hitherto, switching between AD packages in the Julia Language required end-users to familiarize themselves with the user-facing API of the respective packages. Furthermore, implementing a new, usable AD package required AD package developers to write boilerplate code to define convenience API functions for end-users. As a response to these issues, we present AbstractDifferentiation.jl for the automatized generation of an extensive, unified, user-facing API for any AD package. By splitting the complexity between AD users and AD developers, AD package developers only need to implement one or two primitive definitions to support various utilities for AD users like Jacobians, Hessians and lazy product operators from native primitives such as pullbacks or pushforwards, thus removing tedious -- but so far inevitable -- boilerplate code, and enabling the easy switching and composing between AD implementations for end-users. 
\end{abstract}

\section{Introduction} \label{sec:intro}

Differentiable programming (\DP), i.e., the ability to differentiate general computer program structures, has enabled the efficient combination of existing packages for scientific computation and machine learning~\citep{raissi2019physics, rackauckas2020generalized,de2018end}. Black-box machine learning approaches are flexible but require a large amount of data. Incorporating scientific knowledge about the structure of a problem via \DP~reduces the amount of data needed. It allows the learning task to be simplified, for example, by focusing on learning only the parts of the model that are missing~\citep{rackauckas2020universal,dandekar2020machine}.  There are already many examples where such differentiable frameworks have provided performance and accuracy advantages over black-box approaches to machine learning, including but not limited to protein-folding~\citep{AlQuraishi265231,ingraham2018learning}, fluid dynamics~\citep{schenck2018spnets}, robotics~\citep{schenck2018spnets}, and quantum control~\citep{schaefer2020,schaefer2021}.

\DP~is (commonly) realized by automatic differentiation (AD),  a family of techniques to efficiently and accurately differentiate numeric functions expressed as computer programs. Generally, besides forward- and reverse-mode AD, the two main AD branches, many software implementations with different pros and cons exist. Some AD software implementations work at a lower level code representation, possibly mixing in LLVM-level compiler passes, to fully optimize scalar operations~\citep{revels2016forward,enzyme2020} while others perform transformations at a higher level to keep linear algebra operations intact for optimal usage of BLAS primitives~\citep{innes19,paszke2017automatic}. The goal is to make the best choice of AD system in every part of the program without requiring users to extensively contort their code to the differing APIs. 

The AD landscape of the Julia programming language is developed in a manner in which composability between the AD systems is possible. While many automatic differentiation systems require specific formulations of the code, for example PyTorch using an alternative implementation of the NumPy API known as torch.numpy~\citep{paszke2017automatic} with torch.tensor and similarly for Jax with jax.numpy~\citep{jax2018github} each differing from the original NumPy~\citep{oliphant2006guide} API in subtle ways with different numerical properties, the Julia AD systems generally act directly on the standard Julia syntax, with its standard library, array implementation, its standard GPU acceleration tools~\citep{besard2018juliagpu}, and more. This has previously been shown to allow packages in Julia which were developed without knowledge of AD systems to be fully differentiable without modification by multiple different tools~\citep{rackauckas2020generalized}. Furthermore, Julia has a common ground on which differentiation rules are defined, ChainRules.jl~\citep{chainrules2021zenodo}, which is shared amongst the AD packages. This empowers the idea of a ``glue AD'' system~\citep{rackauckas2020glue} where software library authors define ChainRules overloads to add domain insight into the automatic differentiation process without tying to one particular AD system. 

However, switching from one backend to another on the user side can still be tedious because the user has to learn and adapt the code towards the user-facing API of the new AD package. Similarly, for the author of the AD package defining an extensive API supporting every possible differentiation use case requires a lot of boilerplate code, e.g. to define the Jacobian function, Jacobian-vector product, Hessian, Hessian-vector product, etc. Defining all of these functions for each AD implementation is tedious and unnecessary since the relationship between these functions is abstract and not implementation-specific. Therefore, while in theory switching between AD systems can be trivially done, in practice the competing APIs of the various AD mechanisms has limited its use throughout the language's ecosystem.

The Julia Language~\citep{bezanson2012julia} has over a dozen automatic differentiation packages~\citep{juliadiff2021}. Different packages have different user interfaces and offer different tradeoffs. Popular systems include:

\begin{enumerate}
    \item ForwardDiff.jl~\citep{revels2016forward}, an operator-overloading-based, forward-mode AD implementation, with many years of extensive use and thus very high reliability
    \item ReverseDiff.jl~\citep{reversediff}, an operator-overloading-based, reverse-mode AD implementation, featuring several tape-based optimizations
    \item Zygote.jl~\citep{innes19}, a reverse-mode AD implementation that does source code transformation to generate the derivative's code from the function's code, operating at the level of Julia's intermediate representation. Zygote is therefore able to handle arbitrary Julia code but is unable to handle mutation.
    \item Enzyme.jl~\citep{enzyme2020}, a reverse-mode AD implementation that runs by source code transformation at the LLVM level, with excellent performance on scalar operations, but at present lesser performance on large matrix operations.
    \item Diffractor.jl~\citep{diffractor}, a new source-to-source AD package promising high performance on both scalar and vector/tensor code
\end{enumerate}
A more detailed summary of the strengths and limitations of different AD packages is given in Appendix~\ref{sec:ad_packages}. 

Each of these  AD systems (and each of the many others) has its own unique set of advantages and disadvantages. Additionally, all of them only define API functions for a subset of all the possible differentiation use cases, often requiring users to do package-specific implementations of quantities like Jacobian-vector product or Hessian-vector product when needed. Beside the existing stable AD implementations, any new implementation may or may not be mature enough to handle perturbation confusion properly~\citep{siskind2005perturbation, manzyuk2019perturbation} which prevents one from doing general, higher-order AD correctly. A simple workaround is to compose various AD packages for each level of differentiation, further giving rise to applications where changing between AD mechanisms is increasingly common.

As AD systems have different pros and cons, a software author will want to change AD systems depending on the problem and available hardware resources, see Appendix~\ref{sec:AD_perf}. 
However, this is more challenging than it might seem.
Changing AD systems results in forking the code, even though the nominal value of the software using the AD remains the same.
To give some examples: Flux.jl\footnote{\url{https://github.com/FluxML/Flux.jl}} changed from using Tracker.jl\footnote{\url{https://github.com/FluxML/Tracker.jl}} to Zygote.jl~\citep{innes19}.
This resulted in a fork being created, viz. TrackerFlux.jl\footnote{\url{https://github.com/AStupidBear/TrackerFlux.jl}}, for those who want to use the old AD system -- even though conceptually Flux is a Neural Network library that should be abstracted away from the AD.
PyMC4 was created as an attempt to move from Theano~\citep{theano2016}, as used in PyMC3~\citep{PyMC3Salvatier2016}, to using TensorFlow~\citep{tensorflow2015}.
This attempt was eventually abandoned, in favor of keeping Theano but adding a Jax~\citep{jax2018github} backend~\citep{pymc4}.
Not only did the code need to be forked, but the  overall attempt was not successful. Admittedly, this was a particularly complex case beyond just AD, with TensorFlow and Theano being more general computational frameworks with AD as just one feature.
The work we present here aims to ensure that changing the AD system is accessible by providing consistent abstractions that the author of the \DP~ algorithm implementation can use.

A similar but more complex problem was solved by the MathOptInterface.jl~\citep{legat2020mathoptinterface}.
MathOptInterface.jl provides common abstractions across constrained mathematical optimizers such as IPOPT~\citep{ipopt}, Cbc~\citep{cbc}, and Gurobi~\citep{gurobi}.
It in turn is used by mathematical optimization frameworks including JuMP~\citep{JuMP} and Convex.jl~\citep{convexjl}.
Each of the different mathematical optimizers has their own very unique internal set of abstractions, but MathOptInterface.jl exposes them all in the same way.
An additional complication is that each supports different kinds of problems and so this too must be exposed.
Further still, for some classes of problems they can be re-expressed as a different kind through a mathematical transformation, MathOptInterface exposes this through an extensible system of so-called "bridges", that will automatically perform these reformulations.
This system is considerably more complicated than our setting as every AD system can perform all the operations, to varying degrees of efficiency.
The MathOptInterface system has proven very successful, which supports the idea that this kind of abstraction is valuable and can be practically realized.

In light of the above, the authors believe it is necessary to have a backend-agnostic interface to provide objects like the function value, its gradient, Hessian, etc. as well as combining AD implementations together for higher-order AD. Such an interface can help us avoid a combinatorial explosion of code when supporting every differentiation package in Julia in every piece of software requiring gradients and/or Hessians. This is especially important for higher-order derivatives because one can combine any two differentiation backends to create a new higher-order backend. More generally for a $k^{th}$ order derivative, the amount of code required to support $n$ differentiation packages in $m$ \DP~algorithm implementations is $O(m \times n^k)$.

In this paper, we present AbstractDifferentiation.jl~\citep{AbstractDifferentiation}, a package that:
\begin{itemize}
    \item Defines an abstract, extensive API for differentiation in Julia enabling the development of algorithms requiring first and higher-order derivatives in an AD-implementation-agnostic way using a single, unified interface reducing the code complexity from $O(m \times n^k)$ to $O(m + n)$.
    \item Automatically defines most of the extensive user-facing API for any new AD package from just one or two primitive API function definitions, thus making it easier for the AD package developer to support every possible use case without a great deal of boilerplate code.
\end{itemize}

\section{Levels of abstraction in Julia's AD ecosystem}

\begin{figure}[h!]
    \centering
    \includegraphics[width=1\linewidth,angle=0]{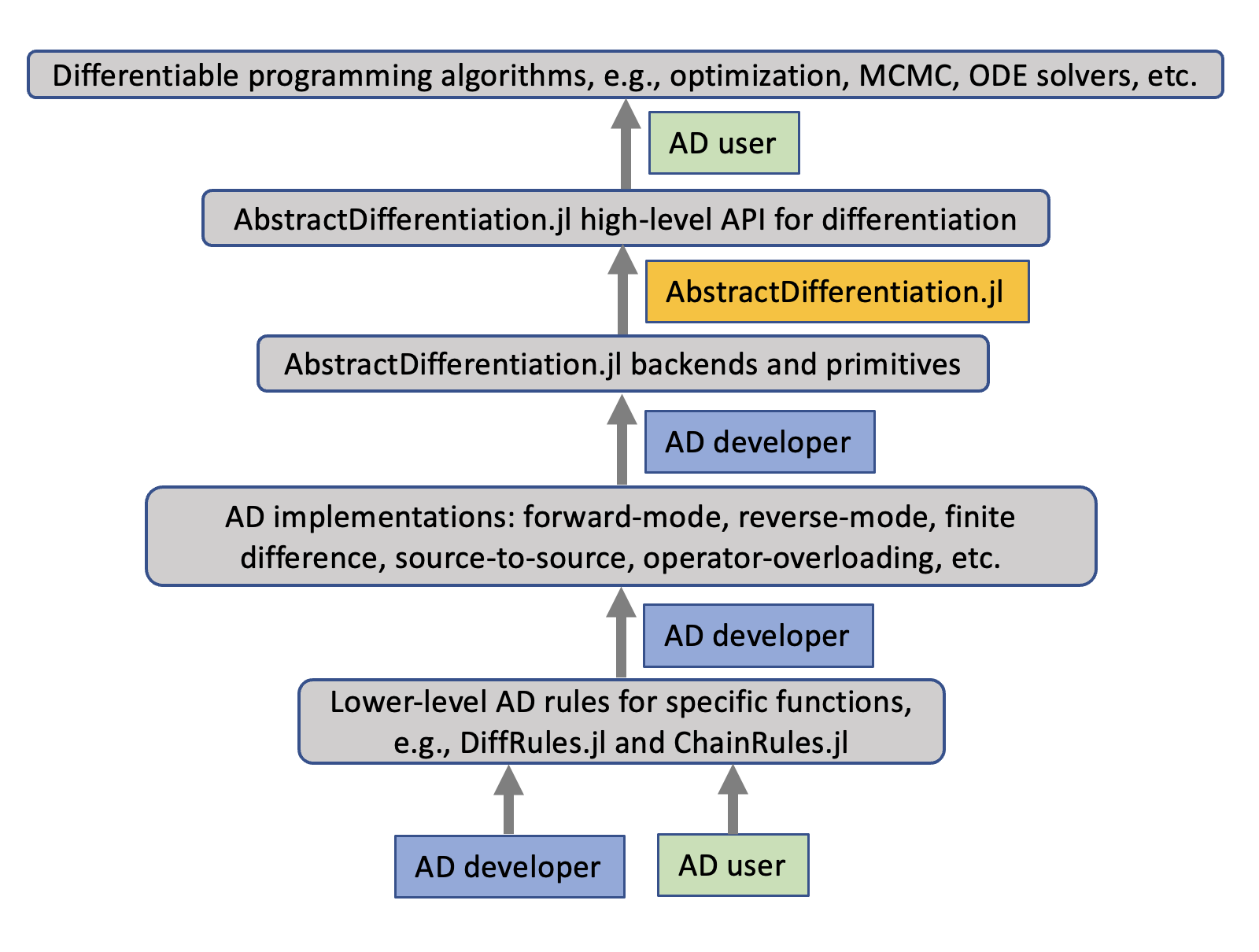}
    \caption{The levels of abstraction in Julia's AD ecosystem.}
    \label{fig:1}
\end{figure}

In Figure \ref{fig:1}, an overview of the levels of abstraction in Julia's AD ecosystem with AbstractDifferentiation.jl is presented. At the bottom level, we have libraries of differentiation rules (DiffRules.jl and ChainRules.jl) for specific functions. These rules are either defined by AD developers for basic Julia constructs, or by AD users for specific user-defined functions with known analytic derivatives.

Sitting on top of the library of rules are all the AD package implementations. At this level, numerous design decisions and optimizations can be made giving a variety of different AD package implementations with different tradeoffs. Each AD package developer will then define a minimal set of primitives and a backend type extending AbstractDifferentiation.jl. These minimal definitions then enable AbstractDifferentiation.jl to automatically define an extensive set of user-facing API functions for AD users to use, e.g. derivative, Jacobian, Hessian, Jacobian-vector product, Hessian-vector product, etc.

At the top level, AD users can then use the relevant part of the \AD~API to implement algorithms requiring \DP. With this abstraction design, the amount of code needed to support all of $n$ AD packages in $m$ algorithms requiring $k^{th}$ order derivatives is only $O(m + n)$, a significant reduction from the $O(m \times n^k)$ without \AD. Additionally, the AD users and developers do not need to add unnecessary boilerplate code to extend an AD package's API anymore, since \AD~automatically does this for them.

\section{API description}

\textbf{Installation and loading}
\AD~is a registered Julia package and can be installed by the \href{https://docs.julialang.org/en/v1/stdlib/Pkg/}{Julia package manager}. The package can be loaded by

\begin{minted}{julia}
# alternatively: import AbstractDifferentiation as AD
using AbstractDifferentiation
\end{minted}

Note that \AD~exports ``AD" as an alias for the AbstractDifferentiation module. This alias allows us to conveniently access names within \AD~via AD instead of typing the full package name.

\subsection{Backends and primitives}

\textbf{Forward-mode, reverse-mode, and finite-difference backends}
All functionalities in \AD~are implemented based on an ab::AbstractBackend type. 
An AD package developer (or the AD user if necessary) first constructs a backend instance that subtypes ab::AbstractForwardMode, ab::AbstractReverseMode, or ab::AbstractFiniteDifference,  which are themselves subtypes of ab::AbstractBackend. For example, backends that support ForwardDiff.jl or Zygote.jl are defined as follows:

\begin{minted}{julia}
## ForwardDiff
struct ForwardDiffBackend <: AD.AbstractForwardMode end
const forwarddiff_backend = ForwardDiffBackend()

## Zygote
struct ZygoteBackend <: AD.AbstractReverseMode end
const zygote_backend = ZygoteBackend()
\end{minted}

By adding fields to the backend struct, we can control configurations of the differentiation package such as chunk sizes, compilation flags, or method choices. To use a finite differencing method at a central grid of 5 points as implemented in the FiniteDifferences.jl package, we write:
\begin{minted}{julia}
## FiniteDifferences
struct FDMBackend{A} <: AD.AbstractFiniteDifference
    alg::A
end
# 1 denotes the order of the derivative to estimate.
FDMBackend() = FDMBackend(central_fdm(5, 1)) 
\end{minted}

\textbf{Higher-order backends}
To compute higher-order derivatives, it may be desirable to combine different backends. 
We provide \verb AD.HigherOrderBackend  to implement higher-order backends. 
Let \verb ab_f  be a forward-mode automatic differentiation backend and let \verb ab_r  be a reverse-mode automatic differentiation backend. To construct a higher-order backend for doing forward-over-reverse-mode automatic differentiation, one defines  \verb AD.HigherOrderBackend((ab_f,ab_r)) . Analogously, higher-order backend for doing reverse-over-forward-mode automatic differentiation is constructed via \verb AD.HigherOrderBackend((ab_r,ab_f)) .

\textbf{Jacobian, pushforward, and pullback as primitive operation}
In addition to the definition of a backend, the AD package developer needs to define one of the following primitive operations: 

\begin{minted}{julia}
AD.@primitive function jacobian(ab::backend, f, xs...)
    return ..
end
AD.@primitive function pushforward_function(ab::backend, f, xs...)
    return ..
end
AD.@primitive function pullback_function(ab::backend, f, xs...)
    return ..
end
\end{minted}
\AD~then generates the other two primitive functions.
For instance, a source-to-source reverse-mode AD package developer can specify only \verb AD.pullback_function  as the native primitive operation.  

\begin{minted}{julia}
## Zygote is source-to-source reverse-mode
AD.@primitive function pullback_function(ab::ZygoteBackend, f, xs...)
    return function (vs)
        # Supports only single output
        _, back = Zygote.pullback(f, xs...)
        if vs isa AbstractVector
            return back(vs)
        else
            # vs isa Tuple
            @assert length(vs) == 1
            return back(vs[1])
        end
    end
end
\end{minted}

In the case of operator overloading AD implementations, we require additionally the definition of \verb AD.primal_value  returning the primal value of the forward pass. 

\subsection{Automatically provided functions}
\label{subsec:functions}

After these preparatory steps, \AD~automatically defines various functions for AD users making use of the primitives defined. Some of the most important API functions provided are presented in the following. We refer the reader to the package documentation for further details~\citep{AbstractDifferentiation}.

\textbf{Derivative, gradient, jacobian, hessian}
\begin{minted}{julia}
ds = AD.derivative(ab::AD.AbstractBackend, f, xs::Number...)
gs = AD.gradient(ab::AD.AbstractBackend, f, xs...)
js = AD.jacobian(ab::AD.AbstractBackend, f, xs...)
h = AD.hessian(ab::AD.AbstractBackend, f, x)
\end{minted}

\textbf{Value and derivative, gradient, jacobian, hessian}

\begin{minted}{julia}
v, ds = AD.value_and_derivative(ab::AD.AbstractBackend, f, xs::Number...)
v, gs = AD.value_and_gradient(ab::AD.AbstractBackend, f, xs...)
v, js = AD.value_and_jacobian(ab::AD.AbstractBackend, f, xs...)
v, h = AD.value_and_hessian(ab::AD.AbstractBackend, f, x)
v, g, h = AD.value_gradient_and_hessian(ab::AD.AbstractBackend, f, x)
\end{minted}

\textbf{Lazy operators}

Finally, we implemented lazy versions of the derivative, gradient, Jacobian, and Hessian,
\begin{minted}{julia}
ld = lazy_derivative(ab::AbstractBackend, f, xs::Number...)
lg = lazy_gradient(ab::AbstractBackend, f, xs...)
lj = lazy_jacobian(ab::AbstractBackend, f, xs...)
lh = lazy_hessian(ab::AbstractBackend, f, x)
\end{minted}
which are of interest in iterative solvers. For example, we compute a vector-Jacobian product by multiplying a single (transposed) vector, or a tuple of an appropriate shape, from the left to the lazy Jacobian operator.

\section{\DP~ use cases and an example}

Many numerical algorithms require the computation of constructs such as the ones described in Section~\ref{subsec:functions}. Table~\ref{Tab:2} presents a rough summary linking some of the most widely adopted routines across different domains to the quantities used in the respective iterative update steps.
As an example, we present a (simple, non-optimized) backend-agnostic implementation of the Gauss-Newton algorithm to solve non-linear least squares problems in Appendix~\ref{sec:Gauss_Newton}.

We also expect \AD~to be specifically handy for (future) AD package like Diffractor.jl or Enzyme.jl where computing constructs like Jacobians or Hessians is technically possible but not yet part of the public API due to abstractions or naming conventions made in the package. 

\begin{table*}[h!]
	\centering
		\begin{tabular}{@{}rcr@{}}\toprule
			\multicolumn{1}{r}{algorithms} & \phantom{abc}& \multicolumn{1}{r}{required quantities} \\
			\midrule
			\textbf{root finding}&& \\
			Newton–Raphson &&  Jacobian\\
			Jacobian-Free Newton Krylov && Jacobian-vector products\\
			\midrule
			\textbf{optimization}&& \\
			ADAM && gradient \\
			Newton && gradient, Hessian \\
			Levenberg–Marquardt && Jacobian \\
			Gauss-Newton &&  Jacobian\\
			\midrule
			\textbf{differential equations}&& \\
			stiff ODE solvers && Jacobian \\
			stiff ODE Jacobian-free solvers && Jacobian-vector products \\
			forward sensitivity methods &&  Jacobian-vector products\\
			adjoint sensitivity methods &&  vector-Jacobian product\\
			\bottomrule
	\end{tabular}
	\caption{Domain-specific algorithms requiring derivatives, gradients, Jacobians, Hessians, vector-Jacobian products, Jacobian-vector products commonly computed by AD packages.}
	\label{Tab:2}
\end{table*}

\section{Summary \& Future work}

The ability to straightforwardly combine different packages in one workflow is one of the most versatile and key features of Julia. Switching between different AD packages and combining them for higher-order derivatives is a useful feature to have when selecting the best AD implementation for a specific application. We have presented the \AD~package which makes this switching and combining of AD implementations as painless as possible for end-users while also reducing the amount of necessary boilerplate code per AD package to support all differentiation use cases.

In the future, we aim to support \AD~ in all of the AD packages in Julia and remove lots of boilerplate code from popular Julia packages (e.g. in the SciML and TuringLang organizations) that heavily employ AD.

\section{Acknowledgments}

This material is based upon work supported by the National Science Foundation under grant no. OAC-1835443, grant no. SII-2029670, grant no. ECCS-2029670, grant no. OAC-2103804, and grant no. PHY-2021825. We also gratefully acknowledge the U.S. Agency for International Development through Penn State for grant no. S002283-USAID. The information, data, or work presented herein was funded in part by the Advanced Research Projects Agency-Energy (ARPA-E), U.S. Department of Energy, under Award Number DE-AR0001211 and DE-AR0001222. We also gratefully acknowledge the U.S. Agency for International Development through Penn State for grant no. S002283-USAID. The views and opinions of authors expressed herein do not necessarily state or reflect those of the United States Government or any agency thereof. This material was supported by The Research Council of Norway and Equinor ASA through Research Council project "308817 - Digital wells for optimal production and drainage". Research was sponsored by the United States Air Force Research Laboratory and the United States Air Force Artificial Intelligence Accelerator and was accomplished under Cooperative Agreement Number FA8750-19-2-1000. The views and conclusions contained in this document are those of the authors and should not be interpreted as representing the official policies, either expressed or implied, of the United States Air Force or the U.S. Government. The U.S. Government is authorized to reproduce and distribute reprints for Government purposes notwithstanding any copyright notation herein.


\bibliographystyle{abbrvnat}
\bibliography{refs}

\newpage

\appendix

\section{Summary of the current state of AD packages in Julia as of September 2021} \label{sec:ad_packages}

\begin{table}[H]
\centering \caption{This table summarizes the current state of popular Julia AD packages in September 2021. ``Scalar" refers to scalar operations support including defining custom rules for scalar-valued functions of scalars. "Vector/tensor" refers to native vector/tensor support as a construct including the ability to define custom differentiation rules for vector/tensor-valued functions and/or functions of vectors/tensors. Similarly, ``First class struct support" refers to the native support of Julia structs as a construct including the ability to define custom differentiation rules for struct-valued functions or functions of structs. ``GPU" refers to the ability to differentiate functions of or returning GPU arrays. "GC" refers to supporting functions that call the Julia garbage collector. ``Mutation" refers to the ability to support mutating arrays and structs. ``Runtime branches" refers to the ability to support ``piece-wise" functions with control flow such that the path that the function takes and ultimately the structure of the mathematical function differentiated depends on the values of the inputs to the function. 
``Maturity" refers to a subjective measure of how mature each package is in the eyes of the community as well as the feature maturity of the package.}
\begin{flushleft}
\begin{tabular}{@{} p{2.6cm} p{0.8cm} p{1.3cm} p{2.4cm} p{0.6cm} p{0.6cm} p{1.5cm} p{1.4cm} p{1.4cm} @{}}\toprule
  Package & Scalar & \makecell{Vector\\/ tensor} & \makecell{First class\\struct support} & GPU & GC & Mutation & \makecell{Runtime\\branches} & Maturity \\
 \midrule
 ForwardDiff & \hfil \cmark  & \hfil \xmark & \hfil \xmark & \hfil \cmark & \hfil \cmark & \hfil \cmark & \hfil \cmark & very high \\
 \midrule
 ReverseDiff & \hfil slow  & \hfil \cmark & \hfil \xmark & \hfil \xmark & \hfil \cmark & \hfil limited & \hfil \cmark & high \\
 \midrule
 \makecell{ReverseDiff with \\ compiled tape} & \hfil \cmark & \hfil \cmark & \hfil \xmark & \hfil \xmark & \hfil \cmark & \hfil limited & \hfil \xmark & high \\
 \midrule
 Tracker & \hfil slow & \hfil \cmark & \hfil \xmark & \hfil \cmark & \hfil \cmark & \hfil limited & \hfil \cmark & high \\
 \midrule
 Zygote & \hfil slow & \hfil \cmark & \hfil \cmark & \hfil \cmark & \hfil \cmark & \hfil \xmark & \hfil \cmark & high \\
 \midrule
 Enzyme & \hfil \cmark & \hfil limited & \hfil wip & \hfil wip & \hfil wip & \hfil \cmark & \hfil \cmark & low \\
 \midrule
 Diffractor & \hfil \cmark & \hfil \cmark & \hfil \cmark & \hfil \cmark & \hfil \cmark & \hfil \xmark & \hfil \cmark & low \\
 \bottomrule
\end{tabular}
\end{flushleft}
\label{tab:ad_packages}
\end{table}

Table \ref{tab:ad_packages} summarizes the current state of the most popular AD packages in the Julia ecosystem as of the time of the writing of this paper.

\newpage

\section{AD performance can be problem-specific}
\label{sec:AD_perf}

It is well know that for a function $f:\mathbb{R}^n \rightarrow \mathbb{R}^m$ with $n$
independent input variables and $m$ dependent output variables, forward-mode AD is preferred to build the Jacobian when $m\gg n$ while reverse-mode AD is preferred when $n\gg m$, i.e. as one increases the number of inputs within the same problem, reverse-mode AD mode will eventually overtake forward-mode AD in performance. This has been investigated in depth for differential equations when applied to models relevant to biopharmacology, alongside various adjoint options~\citep{rackauckas2018comparison}. This work shows that on sufficiently small ODEs (<100 ODEs + parameters), forward-mode AD via ForwardDiff.jl is the most efficient method comparing against analytical techniques and adjoint techniques using Tracker.jl, Enzyme.jl, and ReverseDiff.jl. When the size of the ODEs+parameters is increased in a stiff partial differential equation, it was shown that Enzyme.jl vector-Jacobian products mixed with a specific adjoint method was the most efficient, outperforming the ForwardDiff.jl techniques long with ReverseDiff.jl and Tracker.jl. 

In what follows, we demonstrate on two additional examples that the choice of the specific reverse-mode AD package may also significantly impact the performance~\citep{srajer2018benchmark}. These examples show ReverseDiff.jl in compiled mode outperforming Enzyme.jl under certain circumstances. However, as ReverseDiff.jl is not compatible with GPUs and was shown to not be performance competitive on other potential equations in scientific computing applications, which allows Zygote.jl and Tracker.jl to be the most efficient. Together this shows in one application that 5 AD systems could potentially be the optimal choice depending on user inputs into the package code. This establishes that the optimal choice of AD mechanism can be rather complex for users and package developers, and thus decreasing the cost of performing such benchmarks is of value to many scientists.

\textbf{Example 1: Lotka–Volterra model} 

\begin{figure}[h!]
    \centering
    \includegraphics[width=1\linewidth,angle=0]{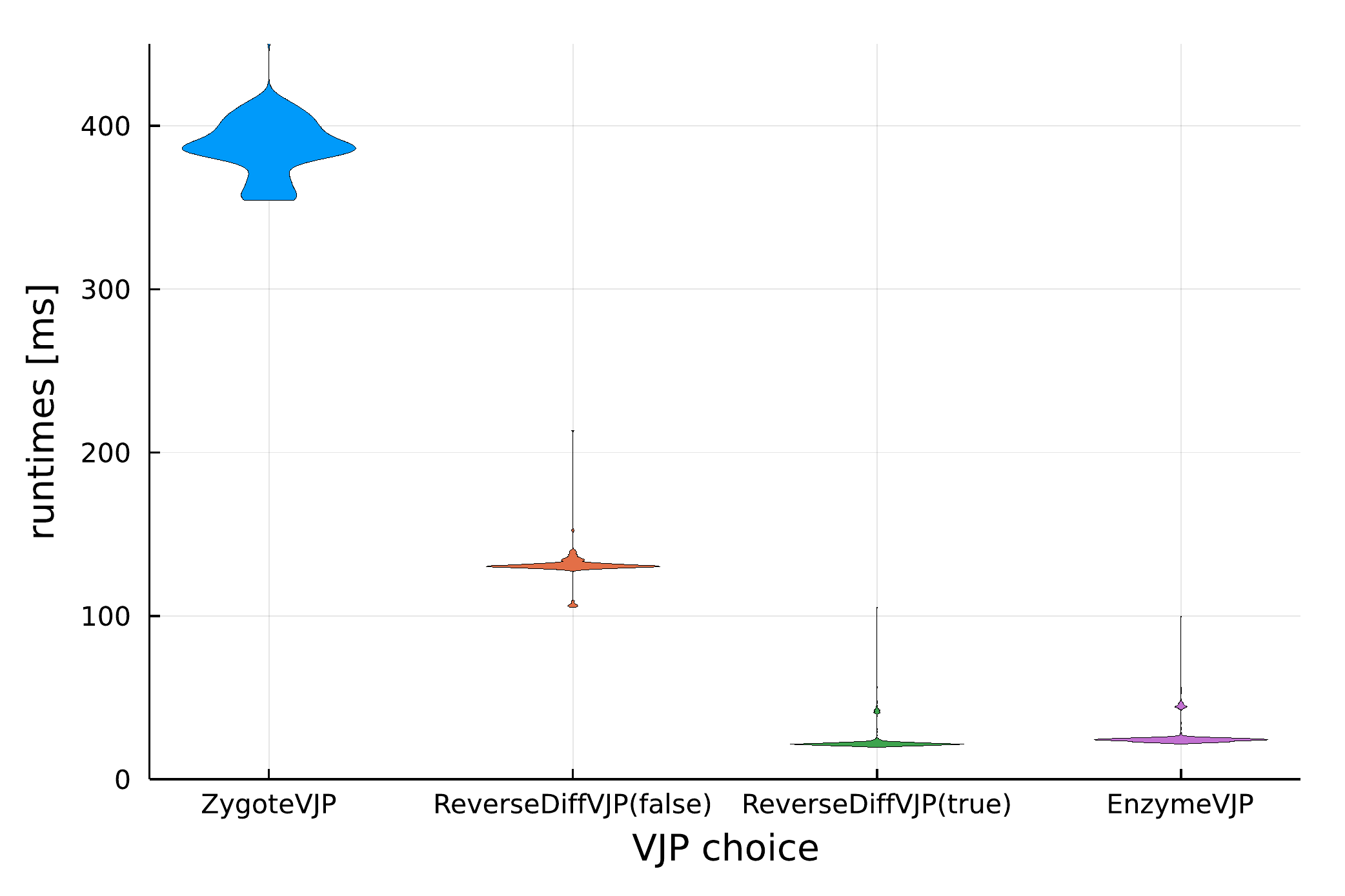}
    \caption{Benchmark 1: Lotka–Volterra model. In all cases, we use a checkpointed interpolating adjoint method~\citep{rackauckas2020universal} to compute the local sensitivities. `false' and `true' indicate if the tape in ReverseDiff.jl is precompiled.}
    \label{fig:2}
\end{figure}
Suppose that we have an instantaneous objective
\begin{align}
    l(x(t),y(t)) = x(t) + y(t) 
\end{align}
for the Lotka–Volterra model
\begin{align}
    \dot{x} &= \alpha x - \beta x y\,,\\
    \dot{y} &= \gamma x y - \delta y\,,
\end{align}
with initial conditions $x(t=0)=1$ and $y(t=0) = 1$. Let $\xi$ denote any of the parameters $\alpha,\beta,\gamma,\delta$. We are interested in the sensitivities $\frac{\partial }{\partial \xi} \sum_i l(x(t_i),y(t_i))$ with respect to an equally spaced time grid between 0 and 10 with a grid spacing of 0.1.

Figure~\ref{fig:2} shows a violin plot for the runtimes for four choices of the internally used AD system.  This demonstrates that the vector-Jacobian products which use static compilation of the ODE function, ReverseDiff.jl with compilation enabled and Enzyme.jl, vastly outperform the other choices for small ODEs with a lot of scalar indexing, which is a common feature in many nonlinear physical and biochemical models. Note that all adjoint techniques were shown to be outperformed by ForwardDiff.jl on this example elsewhere~\citep{rackauckas2018comparison}, but this example still confirms that in many scalar indexing cases the Zygote.jl system can perform rather poorly.

\textbf{Example 2: Neural ODE} 

\begin{figure}[h!]
    \centering
    \includegraphics[width=1\linewidth,angle=0]{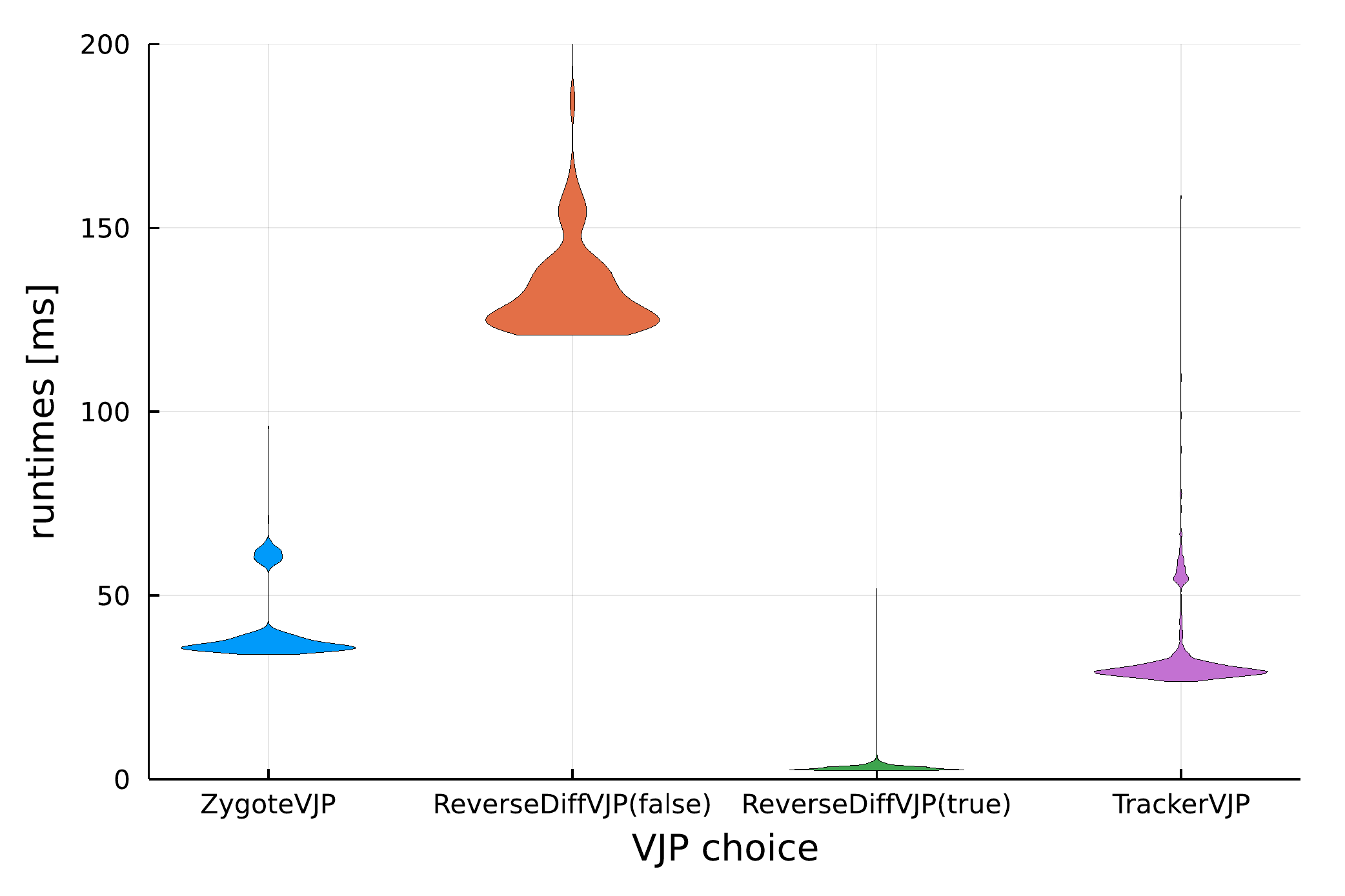}
    \caption{Benchmark 2: Spiral Neural ODE model. In all cases, we use a checkpointed interpolating adjoint method~\citep{rackauckas2020universal} to compute the local sensitivities. `false' and `true' indicate if the tape in ReverseDiff.jl is precompiled.}
    \label{fig:3}
\end{figure}

This example is the Spiral Neural ODE chosen from the Neural Ordinary Differential Equations manuscript~\citep{chen2018neural}. It is an ODE defined as a neural network applied to the cubed states of the system:
\begin{align}
    \dot{u} &= \text{NN}(u^3)
\end{align}
where $\text{NN}(u)$ is a multilayer perceptron with one hidden layer of size 50 and a $\tanh$ activation function, and $u \in \mathbb{R}^2$. Figure~\ref{fig:3} shows a violin plot for the runtimes for four choices of the internally used AD system. The results show that for direct differentiation on CPUs, ReverseDiffVJP with a compiled tape is the most efficient method. However, this has many caveats. One caveat is that ReverseDiff.jl's tape-compiled form is only applicable if the code has no branching, and thus would be incompatible with activation functions like relu. 

Additionally, by testing over various sizes of hidden layers, we established that a RTX 2080 Super GPU outperformed a Ryzen 9 5950x CPU when the hidden layer size reached approximately 7,500 (note the crossover point could potentially be a lot smaller in many scenarios if the neural network is deeper since the first and last layer sizes are 2, matching the dimensionality of $u$). At around this size of neural networks, the Zygote.jl and Tracker.jl strategies on GPUs become more efficient than the one of ReverseDiff.jl which is restricted to CPUs.

These two examples, in addition to the prior research, clearly demonstrate that  the internal AD system must be carefully chosen based on the problem (and hardware resources) at hand. 

\clearpage
\newpage

\section{Implementation of the Gauss-Newton algorithm}
\label{sec:Gauss_Newton}

In this appendix, we use \AD~for the implementation of the Gauss–Newton algorithm for solving nonlinear least squares problems~\citep{schaeferAD}. The Gauss–Newton algorithm iteratively finds the value of the $N$ variables ${\bf{x}}=(x_1,\dots, x_N)$ minimizing the sum of squares of $M$ residuals $(f_1,\dots, f_M)$   

\begin{align}
    S({\bf x}) = \frac{1}{2} \sum_{i=1}^M f_i({\bf x; p})^2.
\end{align}

Starting from an initial guess ${\bf x_0}$  for the minimum, the method runs through the iterations
\begin{align}
{\bf x}^{k+1} = {\bf x}^k - \alpha_k \left(J^T J \right)^{-1} J^T f({\bf x}^k; p),
\end{align}
where the residuals $f({\bf x}^k;p)$ depend on the current step ${\bf x}^k$ and parameters $p$. $J$ is the Jacobian matrix at ${\bf{x}}^k$, and $\alpha_k$ is the step length determined via a line search subroutine.
\begin{minted}{julia}
## Gauss-Newton scheme
function GaussNewton!(xs, x, p backend; maxiter=100)
    for i=1:maxiter
        x = step(x, p, backend)
        push!(xs, x)
    end
    return xs, x
end
function step(x, p, backend, a=1//1)
  x2 = deepcopy(x)
  while !done(x,x2,p) # line-search condition
    # first return value of AD.jacobian is dfdx
    # second return value of AD.jacobian is dfdp
    J = AD.jacobian(backend, f, x, p)[1] 
    d = -inv(J'*J)*J'*f(x,p)
    copyto!(x2,x + a*d)
    a = a//2
  end
  return x2
end
\end{minted}
Switching between different AD systems is then easily accomplished by passing different backends as input to the GaussNewton function. 

\end{document}